# Compact Frequency Standard Based on an Intra-cavity Sample Cold Cesium Atoms

**Stella Torres Müller[1,*], Daniel Varela Magalhães[2], Renato Ferracini Alves[1] and Vanderlei Salvador Bagnato[1]**

[1] *Instituto de Física de São Carlos – USP, Av. Trabalhador São-Carlense 400, 13560-970, São Carlos-SP, Brazil*

[2] *Escola de Engenharia de São Carlos – USP, Av. Trabalhador São-Carlense, 400, 13566-590, São Carlos-SP, Brazil*

[*]*Corresponding author: stella@ifsc.usp.br*

We have demonstrated the possibility for a compact frequency standard based on a sample of cold cesium atoms. In a cylindrical microwave cavity, the atoms are cooled and interrogated during a free expansion and then detected. The operation of this experiment is different from conventional atomic fountains, since all the steps are sequentially performed in the same position of space. In this paper we report the analysis of a Ramsey pattern observed to present a $(47 \pm 5)$ Hz linewidth and a stability of $(5 \pm 0.5) x 10^{-13} \tau^{-1/2}$ for an integration time longer than 100s. Some of the main limitations of the standard are analyzed. This present report demonstrates considerable improvement of our previous work (S. T. Müller et al, J. Opt. Soc. Am. B, v.25, p.909, 2008) where the atoms were in a free space and not inside a microwave cavity.

## I- INTRODUCTION

During the last three decades outstanding developments in the field of atomic frequency standards have been achieved. Today, stabilities in the order of $10^{-16}$ and accuracies that go beyond $10^{-15}$ are possible with laboratory-scale frequency standards employing cold atoms [1].

Most of the high performance time standards require an extremely complex construction and operation, but for certain applications simplified alternatives would be useful. In many cases, smaller and cheaper devices are required, even if they would, in a certain degree, compromise performance. For industrial applications, such as telecommunication lines, mobile telephone networks or internet, calibration of instruments, dissemination of local time references with good quality, etc. commercial atomic clocks must be transportable and easy to use. Commercial atomic clocks are also the basis for many measurement systems such as laser telemetry and global positioning systems [2,3]. The GPS system has proven to be of a high technological importance and is an excellent example of the impact that atomic clocks have in our society. Based on those characteristics, it is difficult to use clocks based on cold atoms for commercial applications, unless simplified devices and methodologies are employed.

The demand for small and practical units has motivated groups to develop Cs beam clocks [4] or optically pumped Rb clocks, offering a huge variety of commercial models. The use of compact high-performance clocks [5,6] is of crucial importance to improve independence and reliability of inertial systems embedded in satellites and submarines.

The aim of this work is to report the construction and first performance of an experimental compact atomic reference, using cold atoms in a device presenting simplicity in two important aspects: construction and operation (compared to the complexity of other systems based on cold atoms as a "fountain"). The use of cold atoms in this work represents a huge advantage over the existing commercial clocks based on atomic beams. Cold atoms naturally allow the achievement of higher precision due to their narrow velocity distribution.

The key idea of our device is to perform in a microwave cavity each stage required by an atomic frequency standard using cold atoms. This is done using atomic cooling techniques, for a vapor inside the cavity and a two pulses method thus achieving the desired compactness of the clock physics package, reducing it to a few liters.

A proof of concept for this idea was published in a recent paper [7], where a free expanding cloud of cold atoms was arranged to interact with a microwave field emitted from an antenna. In that set up [7] a considerable reduction of the fringe contrast was observed and interpreted, and possible improvements were identified. Such a limitation has presently been overcome using a resonator instead of an external antenna, and this is precisely the main technical improvement presented in this work. The final result of this configuration has shown to be surprisingly good.

The first and second part of this publication is dedicated to the description of the experiment and the employed temporal sequence. In the third part, a typical evaluation of the device is presented and dominant frequency shifts are identified.

## II- EXPERIMENTAL SET UP

The atomic sample preparation begins with a magneto-optical trap (MOT) [8] for Cesium atoms operating in the $D_2$ line (852 nm). The configuration is composed of three pairs of counter propagating laser beams orthogonally distributed. In brief, two stabilized diode lasers, mounted in an extended cavity configuration, provide the frequencies necessary to produce trapping, population repumping, and analysis of the clock microwave transition, through fluorescence.

Once a significant number of cold atoms has been captured, a MOT with a density in the range of $10^{10}$ cm$^3$ is obtained, the next phase consists of performing subdoppler cooling with the MOT magnetic field switched off, operating in a molasses configuration. Finally the laser beams are turned off. During the free expansion, the cloud is subjected to a sequence of two microwave pulses, composing the well-known Ramsey method [9]. The microwave pulses are applied in a cylindrical microwave cavity that was sculpted inside the stainless steel vacuum chamber as shown in Fig.1 (a and b).

The whole experiment takes place inside this cavity and the most relevant steps can be seen in Fig.2. The vacuum system is maintained below $10^{-7}$ Pa using an ion pump.

The cavity is used as a microwave resonator for the $^{133}$Cs and it is fed by an intra-cavity antenna coupled to a microwave chain. Obviously, the volume of the cavity is defined by the clock transition $6^2S_{1/2}|F=3, m_F=0\rangle \rightarrow 6^2S_{1/2}|F=4, m_F=0\rangle$, at 9,192,631,770 Hz and couples the TE011 mode [10]. The overall level diagram and used transitions are depicted in Fig.3. The quality factor of the microwave cavity was measured to be close to 1600.

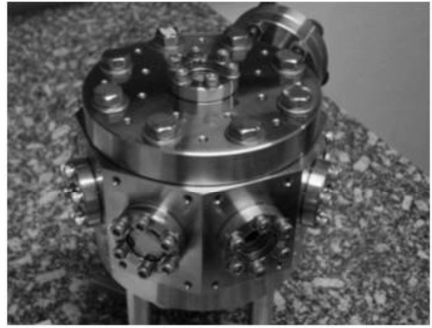

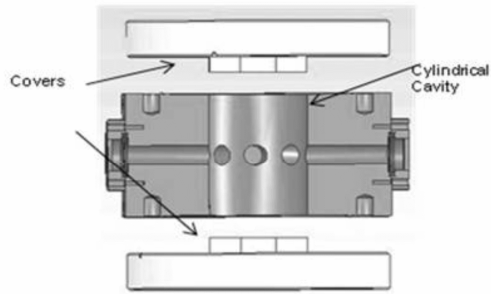

Figure 1- (a) View of the vacuum chamber built for the compact clock project (b) Cutting view of the cylindrical microwave cavity sculpted inside a vacuum chamber, where all interactions take place.

Fluorescence light from the intersecting laser beams region is imaged onto a calibrated detector, located outside the cavity, to perform the detection phase. A set of three pairs of coils allows control of the magnetic field in the atomic cloud region as well as the establishment of a C-field. Due to the high sensitivity of the experiment to the magnetic field, a procedure to obtain a zero field is carried out through careful measurement of the regularity of the free expansion after switching off the trap field.

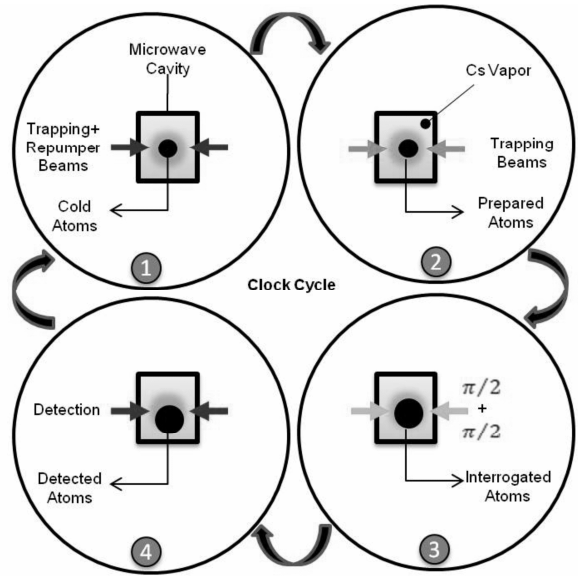

Figure 2 – Temporal working cycle: Each interaction is performed sequentially inside the microwave cavity. (1) Loading in a MOT configuration to confine atoms in the F=4 state, followed by a molasses phase and a frequency sweep to obtain subdoppler cooling of the atoms; (2) The repumping light is switched off and the atoms are optically pumped to F=3 state. After some milliseconds, the main beams are also switched off; (3) A sequence of two microwave pulses is applied during a total interrogation time of 8ms; (4) The main beams are switched on again, tuned to a cycling transition, and the fluorescence of the cloud is acquired by a photodetector.

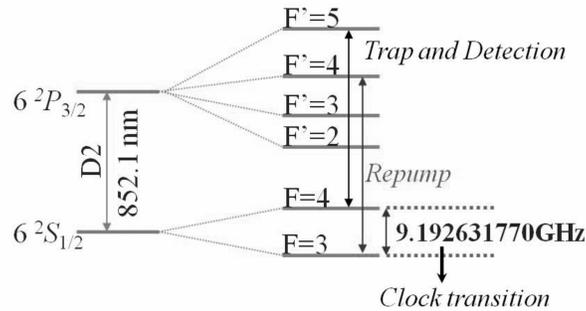

Figure 3- Energy levels for the $D_2$ line of Cs atoms and the transitions used to trap, repump and detect the atoms.

## III- TEMPORAL SEQUENCE OF OPERATION

A typical temporal sequence is divided into 4 parts: cooling, preparation, interrogation and detection. The control signals required to time the operating cycles are presented in Fig.4. They are defined in a LabView program and generated in a data acquisition computer card (PCI-6025E National Instruments) as described in details as follows:

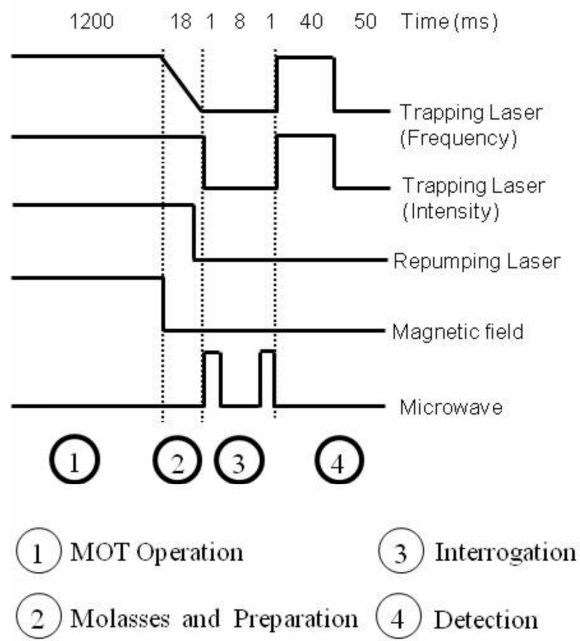

Figure 4 - Control signals of the temporal sequence. The scale is intentionally not preserved to allow better view of the different states.

### *Cooling*

The first part of the cycle consists in loading $10^8$ atoms in a magneto optical trap (MOT) for about 1 s from a cesium vapor. The $6S_{1/2}|F=4\rangle \rightarrow 6P_{3/2}|F'=5\rangle$ cycling transition is used, and repumping is done using the $6S_{1/2}|F=3\rangle \rightarrow 6P_{3/2}|F'=4\rangle$ transition.

*Preparation*

After the MOT phase, an optical molasses configuration is set for a period of 13 ms by turning off the trapping coils. At the end of the molasses cycle, a cloud of atoms at about 10 μK is obtained. Then, in order to prepare the atoms in their ground electronic state $6S_{1/2}|F=3\rangle$, the repumping light is switched off for 5 ms before shutting off the cooling light. During this time optical pumping efficiently transfers the atoms to the required ground state.

*Interrogation*

After the total shutdown of the laser beams, the atoms start a free expansion with a small fall under gravity. During this expansion, the microwave Ramsey Interrogation method [9] of the $6^2S_{1/2}|F=3, m_F=0\rangle \rightarrow 6^2S_{1/2}|F=4, m_F=0\rangle$ clock transition is applied, using two pulses of 1 ms, separated by 8 ms. The microwave chain used was built by F. Walls group (NIST – Boulder – USA) [11,12], and the modulation is produced by an external function generator (DS345 Stanford) synchronized with the chain.

*Detection*

To perform the detection phase, the light beams originally used as a cooling laser are turned back on for 40 ms, and the fluorescence signal of atoms in the $6S_{1/2}|F=4\rangle$ level is detected. The detected population is the one originated from the clock transition due to the interaction with the microwave radiation.

Once the resonance is observed, it can be used as a frequency discriminator, generating an error signal that represents the frequency fluctuations between the atomic resonance and the local oscillator (microwave chain). The frequency correction is sent to the external function generator which provides a nearly 9192631770 Hz signal. Since the microwave chain is phase-locked to a Hydrogen Maser (CH1-75A Kvarz) the error signal is then computed to obtain the frequency stability using the Allan Variance [13].

## IV- RESULTS AND DISCUSSIONS

In parallel with the experiment, we performed numerical simulations of the cloud in free expansion inside the cylindrical microwave cavity. We started by considering the microwave cavity as depicted in Fig.5. The purpose of the simulation is to see the expected Ramsey pattern for our configuration, and to predict any possible external influences.

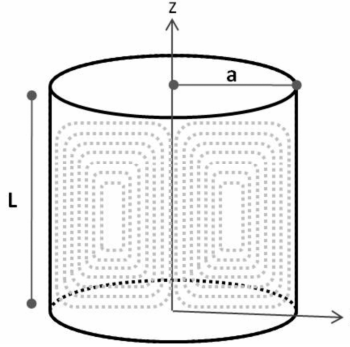

Figure 5 – Drawing of a cylindrical microwave cavity of - radius *a* and length *L*. The dotted lines show the microwave field for the mode $TE_{011}$.

A Labview program was used to simulate the expansion of the atomic cloud inside the microwave cavity taking into account a Boltzmann distribution and the interaction with the microwave field for the mode $TE_{011}$ [9]. The magnetic field in this mode is expressed in cylindrical coordinates as follows:

$$\vec{H}(\vec{r}) = \begin{cases} H_r = \dfrac{\pi a}{x'_{01}} \cos\left(\dfrac{\pi z}{L}\right) J_1\left(\dfrac{x'_{01} r}{a}\right) \\ H_\theta = 0 \\ H_z = \sin\left(\dfrac{\pi z}{L}\right) J_0\left(\dfrac{x'_{01} r}{a}\right) \end{cases} . \quad (1)$$

where *a* is the radius and *L* is the length of the cylinder. $x'_{01} = 3.832$ is the first root of $J'_0(x) = 0$, the derivative of $0^{th}$ order Bessel function of the first kind *J*.

The atomic cloud does not have a uniform density distribution and, therefore, different quantities of atoms interact with the field during the expansion at different amplitudes due to each location along the microwave cavity. Considering a Gaussian atomic density distribution, a shell of radius *R*, as represented in Fig.6, corresponds to a region of a uniform atomic density. The integration of many shells is the main consideration for the simulation.

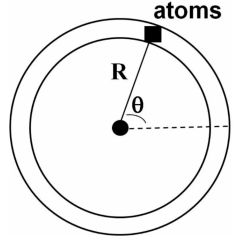

Figure 6 – A diagram showing the angle and the spatial distribution of the atoms used in the simulation.

If *n(R)* is the density distribution, the quantity of atoms between *R* and *R+dR* is:

$$dN = 2\pi R^2 n(R)\sin\theta d\theta dR. \qquad (2)$$

with $n(R) = n_o e^{\frac{-R^2}{W^2}}$, *W* being the Gaussian width representing the cloud expansion and $n_o$ the central peak density.

For this spatial distribution one can calculate the field experienced by each group of atoms at different locations and, therefore, giving different contributions to the signal.

The result of our simulation is shown in Fig.7. The loss of the fringe contrast is attributed to the amplitude difference experienced by the atoms for each microwave pulse, since the cloud of atoms is in free expansion. It should also be noted that the module of the microwave field decreases by 10% compared to the maximum amplitude when the atoms are 6 mm above or below the center of the cavity.

The experimental result obtained in our device is presented in Fig.8 which shows a typical scan of the microwave frequency across the clock resonance. A large variety of investigations of the Ramsey fringes are presented in [14]. Experimental data are averaged over five measurements. The central fringe width is $(47 \pm 5)$ Hz for an operation of an 8 ms Ramsey pulse separation. The contrast is better than 80% (where the contrast is defined as a difference between resonance amplitude and background). The experimental data are in good agreement with the calculated curve (Fig.7), showing that the main physical insights have been well understood. A detail for a scan around the central fringe obtained in Figure 8 is presented separately in Fig.9, together with a theoretical fitting.

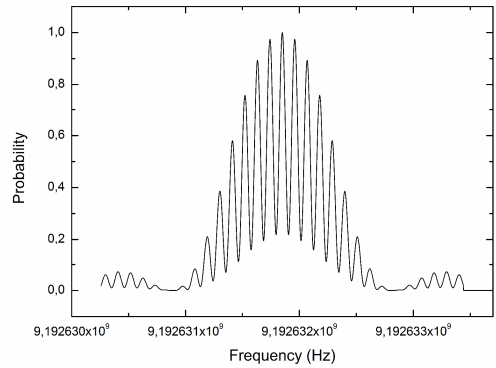

Figure 7 – Simulation of the expected fringes when we consider the relative motion of atoms with the microwave field profile.

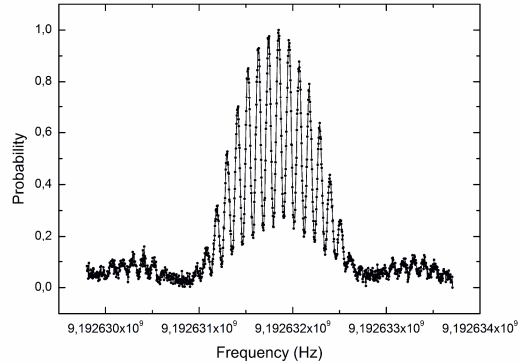

Figure 8 – Observed resonance for the clock transition using Ramsey pulses of 1 ms separated by 8 ms. The FWHM is close to 47 Hz.

The central resonance observed in Fig.9 was used as a clock transition to measure the frequency stability. The microwave chain was phase locked to the 10 MHz output of the Hydrogen Maser. The modulation of the frequency is controlled by a computer, which also registered the introduced correction to keep the interrogation signal at its maximum.

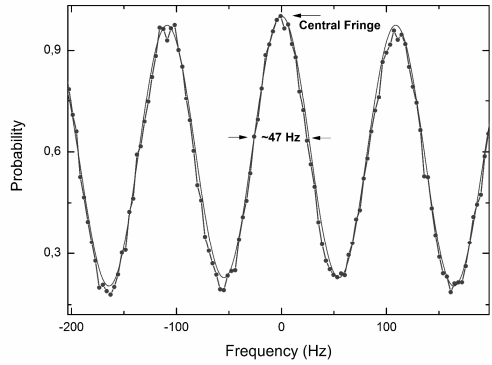

Figure 9 – Scan over the experimental points of the central Ramsey fringes observed in fig. 8, superposed to simulation represented by the continuous line. The contrast is better than 80%.

We observed a stability of $\sigma_y(\tau) = (5 \pm 0.5) \times 10^{-13} \tau^{-1/2}$ after 100 s of integration time, as presented in Fig.10, demonstrating that this compact cold atom clock can reach better performances than the commercial beam clocks, which typically have short time stabilities on the range of $10^{-11}$ - $10^{-12}$. Below 100 s the slope is dominated by $\tau^{-1}$ behavior and we believe that the most significant noise comes from the detection noise. This can be improved using an absorption detection and by cycling the experiment faster. The absorption detection should minimize the loss of atoms in the detection phase and consequently decrease the loading phase, which is the longest time we have in the sequence. This should result in decreasing the dead time.

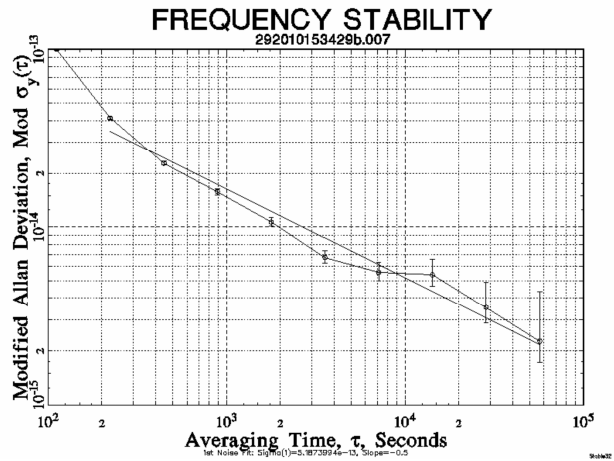

Figure 10 – Allan standard deviation of the system when the central fringe is locked to a Hydrogen maser.

With the system employed here, we obtained a stability two orders better than the result reported in our original "proof of concept" experiment [7,14] and a observed linewidth with a contrast better than 80%.

The huge potential for the use of the cold atom can be realized if we compare the linewidth of both situations. While the Ramsey interrogation method in the present system corresponds to a linewidth of $(47 \pm 5)$ Hz, this is comparable with a linewidth of 26 Hz based on a thermal Cs beam of 4 m long transition region [15]. This comparison shows the technical advantages of a cold atom clock over the traditional beam clock. This is an important feature, since our goal is to benefit from the laser cooling techniques to reach higher stabilities than the commercial thermal Cs beam.

We found comparable results to those obtained by Trémine and collaborators [16,17] in the "Horace" project, an experiment similar to the one here presented, but using a different approach to capture the atoms. The Horace project uses a spherical copper cavity that was optically polished in order to build an isotropic light field to trap the cesium atoms. At the bottom of this device, a time of flight detection region is used to measure the number of atoms by absorption.

In an atomic standard, even with many protection devices, the atoms are disturbed by environmental factors such as temperature, magnetic field, spurious light, and others. There are many interactions and effects that shift the value of the atomic resonance compared to the hypothetical situation of an isolated atom. It is necessary to identify these effects and evaluate their impact on the clock to correct them. We show in Table 1 a list of preliminary systematic effects that affect the operation of our experiment. A more complete evaluation including other shifts shall be done soon.

Table 1 – Preliminary list of systematic effects in a compact atomic clock.

| *Physical Effect* | *Correction* | *Uncertainty* |
|---|---|---|
| Red Shift | $-1.0 \times 10^{-14}$ | $-5.4 \times 10^{-15}$ |
| 2$^{nd}$ order Doppler Shift | $-4.2 \times 10^{-19}$ | $10^{-21}$ |
| 2$^{nd}$ order Zeeman Shift | $4.1 \times 10^{-9}$ | $0.45 \times 10^{-9}$ |
| Black Body Radiation Shift | $-1.84 \times 10^{-14}$ | $-1.0 \times 10^{-16}$ |

Due to the fact that the atomic standard is based on an expanding cloud of atoms, it has no stringent size limitations. One can imagine the possibility of a clock even more compact. In fact, this is the next step of this project, with all essential parts built in a single block.

The compact atomic standard based on cold atoms can be an important contribution to a primary standard of high relevance, and a possible strategic product with a broad range of applications.

## V- CONCLUSIONS

We have described the construction and preliminary performance of a compact cold atom standard using an all-in-one configuration. Some frequency shifts are reported and a stability of $\sigma_y(\tau) = (5 \pm 0.5) x 10^{-13} \tau^{-1/2}$ for a clock transition with a linewidth of $(47 \pm 5)$ Hz was measured. These results show the potential use of clocks with an operation as described here compared to a cesium beam standard. Nevertheless, it is necessary to conduct further investigation in order to improve the clock stability, complete the evaluation of the frequency shifts and evaluate the accuracy. For the next step of our ongoing project we will develop a system containing all the laser sources, microwave source and cavity in a single metallic block. The MOT will also be substituted by a molasses, so avoiding transient magnetic fields.

## ACKNOWLEDGMENT

The authors are pleased to acknowledge A. Clairon, S. Bize and G. Santarelli, from the Observatoire de Paris, for all the fruitful discussions and advices. We acknowledge financial support from FAPESP, CNPq and CAPES. Special acknowledgment is given to the program FAPESP-CNRS.